%% file: main.tex
\documentclass[twocolumn]{aastex63}

\usepackage{amsmath}

\accepted{March 9, 2024}

\submitjournal{ApJ}

\shorttitle{Hawc+ Observations of IRAS 16293-2422}
\shortauthors{Encalada et al.}

\graphicspath{{./}{figures/}}

\begin{document}

\title{Magnetic Fields Observed along the E-W Outflow of IRAS 16293-2422}

\correspondingauthor{Frankie J. Encalada}
\email{fje2@illinois.edu}

\author[0000-0002-3566-6270]{Frankie J. Encalada}
\affiliation{Department of Astronomy, University of Illinois, 1002 West Green St, Urbana, IL 61801, USA}

\author[0000-0002-4540-6587]{Leslie W. Looney}
\affiliation{Department of Astronomy, University of Illinois, 1002 West Green St, Urbana, IL 61801, USA}
\affil{National Radio Astronomy Observatory, 520 Edgemont Rd., Charlottesville, VA 22903 USA}

\author[0000-0003-1288-2656]{Giles Novak}
\affil{Center for Interdisciplinary Exploration and Research in Astronomy (CIERA), Northwestern University, 1800 Sherman Avenue,  
Evanston, IL 60201, USA}
\affil{Department of Physics \& Astronomy, Northwestern University, 2145 Sheridan Road, Evanston, IL 60208, USA}

\author[0000-0001-7474-6874]{Sarah Sadavoy}
\affil{Department for Physics, Engineering Physics and Astrophysics, Queen’s University, Kingston, ON K7L 3N6, Canada}

\author[0000-0002-5216-8062]{Erin G. Cox}
\affil{Center for Interdisciplinary Exploration and Research in Astronomy (CIERA), Northwestern University, 1800 Sherman Avenue,  
Evanston, IL 60201, USA}

\author{Fabio Pereira-Santos}
\affil{Max Planck Institute for Astronomy, K\"onigstuhl 17, 69117 Heidelberg, Germany}

\author[0000-0002-3455-1826]{Dennis Lee}
\affil{Center for Interdisciplinary Exploration and Research in Astronomy (CIERA), Northwestern University, 1800 Sherman Avenue,  
Evanston, IL 60201, USA}
\affil{Department of Physics \& Astronomy, Northwestern University, 2145 Sheridan Road, Evanston, IL 60208, USA}

\author[0000-0003-2118-4999]{Rachel Harrison}
\affiliation{Department of Astronomy, University of Illinois, 1002 West Green St, Urbana, IL 61801, USA}
\affiliation{School of Physics and Astronomy, Monash University, Clayton VIC 3800, Australia}

\author[0000-0002-8557-3582]{Kate Pattle}
\affil{Department of Physics and Astronomy,
University College, Gower St., London WC1E 6BT, UK}

\begin{abstract}

Magnetic fields likely play an important role in the formation of young protostars. Multiscale and multiwavelength dust polarization observations can reveal the inferred magnetic field from scales of the cloud to core to protostar.
We present continuum polarization observations of the young protostellar triple system IRAS 16293-2422 at 89 $\mu$m using HAWC+ on SOFIA.
The inferred magnetic field is very uniform with an average field angle of 89$^\circ\pm$23$^\circ$ (E of N), which is different from the $\sim$170$^\circ$ field morphology seen at 850 $\mu$m at larger scales ($\ga$ 2000 au) with JCMT POL-2 and at 1.3 mm on smaller scales ($\la$300 au) with ALMA.
The HAWC+ magnetic field direction is aligned with the known E-W outflow. 
This alignment difference suggests that the shorter wavelength HAWC+ data is tracing the magnetic field associated with warmer dust likely from the outflow cavity, whereas the longer wavelength data are tracing the bulk magnetic field from cooler dust.
Also, we show in this source the dust emission peak is strongly affected by the observing wavelength. 
The dust continuum peaks closer to source B (northern source) at shorter wavelengths and progressively moves toward the southern A source with increasing wavelength (from 22 $\mu$m to 850 $\mu$m).

\end{abstract}


\section{Introduction} \label{sec:intro}
Magnetic fields are thought to be a critical component of the star formation process on all scales, from the assembly of molecular cloud structure to protostellar accretion \citep[e.g.,][]{pattle2023,tsukamoto2023}.  Nevertheless, the details of how magnetic fields affect star formation are not well observationally constrained.
The magnetic field is difficult to directly observe in most regions. The most common approach is to use dust polarization emission to infer the plane of the sky magnetic field orientation.  This is possible due to radiative alignment torques that tend to align the short axis of elongated dust grains with the magnetic field \citep[e.g.,][]{lazarian2007}. 
As a result, thermal emission of dust grains will be polarized in a direction that is perpendicular to the field. Many studies therefore rotate the polarization angles by 90$^\circ$ to present the inferred plane-of-sky magnetic field morphology.
Although this technique works very well in most star forming regions and spatial scales, when resolving the circumstellar disk other polarization mechanisms, such as scattering, dominate \citep[e.g.,][]{kataoka2015, yang2016,cox2018,harris2018,sadavoy2019}.  This makes it very difficult to measure the magnetic field morphology on small scales using dust polarization observations.

On the other hand, there have been a few surveys of the magnetic field in the inner envelope of protostars comparing the average B-field axis with the outflow axis
\citep{Hull2014,Galametz2018,Zhang2014}. Most recently, \cite{Huang2024} surveyed 61 protostars in Orion (with 56 detections) from the Class 0/I identified disk sources in the VANDAM survey \citep{tobin2022}. They found that $\sim$40\% of the protostars exhibit B-fields that are perpendicular to the source outflows on scales of 400-1000 au, with the remainder of the sources being consistent with random alignment. 

The magnetic field may also play roles in the formation and/or evolution of binaries and multiple systems.
Main sequence stars exhibit a multiplicity rate that increases 
with stellar mass: the mean frequency of stellar companions per primary rises from 0.5 for solar-type main sequence stars to 2.1 for O-type main sequence primaries \citep{moe2017}. 
In addition, the multiplicity fraction is largest in the youngest populations of protostars \citep{tobin2016,encalada2021,tobin2022}, likely decreasing as the systems age.  To better understand binary formation, we must observe the youngest protobinary systems during binary formation. For young binary systems with separations $<$500 au, we expect the formation pathways are likely a combination of disk fragmentation by gravitational instability and turbulent fragmentation with migration, whereas in systems with separations $>$1000~au the binary formation mechanism is likely dominated by turbulent fragmentation \citep[e.g.,][]{adams1989,lee2019,padoan2002,moe2018}. 

IRAS 16293-2422 is a well studied Class 0 protostellar triple system located in the $\rho$-Ophiuchi star forming region inside the dark cloud L1689N \citep{lynd1962} at a distance of 141 pc \citep{dzib2018}. The IRAS 16293 system consists of two close binaries in the southeast (separated by 54 au), sources A1 and A2, \citep{al1989,maureira2020} and a single protostar in the northwest, source B, (separated by 725 au) \citep{al1989,looney2000}. Sources A and B are embedded in a large (6-8 $\times$ 10$^3$ au) envelope \citep[e.g.,][]{schoier2002,crimier2010} with a bridge of material connecting them \citep[e.g.,][]{looney2000,pineda2012}. Although there had been speculation that source B was a more evolved T Tauri star \citep{Stark2004}, observations of an inverse P-Cygni infall profile toward source B \citep{pineda2012} indicate that the differences in the sources are more due to viewing angle, as source B is likely face-on \citep[e.g.][]{Rodriguez2005} whereas the binary sources A1 and A2 are more edge-on \citep{pineda2012}.

With three protostars and their geometry, the system has both complicated envelope and outflow structures.
The envelope has complex chemistry \citep[e.g.,][]{jorgensen2016}, and due to the multiplicity and the connecting bridge, the envelope presents difficulties in modeling the source details \citep{jacobsen2018}.
The system has large and impressive outflows that even at their discovery were known to be multi-lobed and very complicated, likely due to a multiple system \citep[e.g.,][]{fukui1986, al1987,walker1988,mizuno1990}. The multiple outflows, some of which are observed only on large scales while others are observed only on small scales, are well summarized in \cite{vanderwiel2019} and many references within.

On the largest scales, there are two observed outflows: east (blue)-west (red) \citep{fukui1986, al1987,walker1988,mizuno1990, Stark2004} and northeast (red)-southwest (blue) \citep{walker1988,mizuno1990,Stark2004}. However at smaller scales, the northeast-southwest outflow is not detected, implying that the launching engine was  quenched \citep{vanderwiel2019}.  On the other hand, the east-west outflow, which is driven by source A, is clearly detected on smaller scales with interferometers \citep{Girart2014,Yeh2008,vanderwiel2019}.
An additional outflow detected only on the small scale is aligned northwest (blue)-southeast (red) originating slightly north of source A with a blue bowshock in the location of source B \citep{Girart2014,kristensen2013}.
Although there are some observational hints of an outflow from source B \citep[e.g.,][]{Yeh2008}, there are no observations that show clear evidence of outflow emission from source B.

In this paper, we present far-infrared observations (89 $\mu$m) on the scale of $\sim$1000 au toward IRAS 16293-2422.
The observations used the High-Resolution Airborne Wide-band Camera
\citep[HAWC+:][]{vaillancourt2007,harper2018} onboard the Stratospheric Observatory For Infrared Astronomy (SOFIA). We use the 89 $\mu$m dust continuum observations to infer the magnetic field in the envelope of IRAS 16293-2422, hereafter called IRAS 16293. 
The inferred magnetic field morphology is compared to 
850 $\mu$m observation from the POL-2 polarimeter on the James Clerk Maxwell Telescope (JCMT) and to 
1.3 mm observations from the Atacama Large Millimeter/submillimeter Array (ALMA).
In addition, we compare the peak flux location variation with wavelength using archival observations from WISE, Herschel, and SCUBA 2 on the JCMT.

The paper is organized as follows: in Section 2 we present the observations and data reduction, Section 3 covers the results of the observations, Section 4 discusses the results, and Section 5 contains our conclusions.

\section{Observations and Data Reduction} \label{sec:observations}

We observed IRAS 16293 on 2019 July 23rd, between 12:21:13 and 12:25:41 UTC, using HAWC+ at Band C (89 $\mu$m). The observations were part of the Cycle 7 program PID 07\_0147 (PI: Novak). The bandwidth was 16.9 $\mu$m, the angular resolution (full width at half maximum; FWHM) was 7.8$^{\prime\prime}$, and the field of view for the total intensity and polarimetry was 4.2$^{\prime}$ $\times$ 2.7$^{\prime}$ and 2.1$^{\prime}$ $\times$ 2.7$^{\prime}$, respectively. 

The HAWC+ instrument is a far-infrared detector with 5 broadband filters for continuum, a rotating half-wave plate that modulates the incoming polarization, and a grid that orthogonally separates the polarization. Multiple cycles of standard nodding and chopping were used to build an image within a few minutes \citep{hildebrand2000}. The instrument typically nods at 0.1 Hz and chops at 10 Hz. The four dithered positional observations are used to construct a single imaging block, or dither set. Due to a lack of bright visual guide stars in the region, there was a small offset in some of the imaging blocks. To correct this, we shifted the fit map Gaussian peak center to align with the archival map from Herschel's Photodetector Array Camera \& Spectrometer (PACS) at 100 $\mu$m.
No individual observation was offset by more than half a beam.

The raw images are then processed through the HAWC+ data reduction pipeline, which produces science ready continuum and polarization products. The details of the process are summarized in \cite{harper2018}.
Briefly, the pipeline flat-fields the demodulated and usable chopped data, calculates the Stokes I, Q, and U parameters from the combined nodded fluxes, and then corrects the pointing and instrumental polarization. Calibrations are applied to the flux via standard atmospheric opacity models using Neptune planetary observations. Finally, the Stokes maps are combined via standard re-gridding and Gaussian smoothing \citep{houde2007}. 
We smoothed the final maps to increase the signal to noise in the lower surface brightness regions away from the central source, creating a final beam size of 11.7$^{\prime\prime}$. 

We compared our Stokes I, Q, and U maps across the different dither sets using a $\chi^{2}$ analysis. A $\chi^{2}$ is calculated for each pixel, which is then compare with the pipeline error result for that pixel. If the $\chi^{2}$ is larger, then we inflate the pipeline error  \citep[also see][]{Chapman2013,Novak2011}. 
Following \cite{Erin2022}, we
perform the inflation by fitting a parameterized $\chi^{2}$ with Stokes I intensity, allowing us to inflate each pixel based on its continuum brightness.
This was done to handle the uncertainties more carefully in the the brightest regions, where intensity-dependent errors dominate.  

Finally, the percent polarization ($p$) and its error ($\sigma_{p}$) are calculated for each pixel following the HAWC+ handbook \citep{hawcplus_handbook} by

\begin{equation}
p =  100 \sqrt{\left(\frac{Q}{I}\right)^2 + \left(\frac{U}{I}\right)^2}
\end{equation}

and

\begin{equation}
\sigma_{p} = \frac{100}{I} \sqrt{
\begin{aligned}
\frac{1}{(Q^2 + U^2)}
(Q\sigma_{Q}^2 + U\sigma_{U}^2 + 2QU\sigma_{QU})+ \\
\left[\left(\frac{Q}{I}\right)^2 + \left(\frac{U}{I}\right)^2\right]\sigma_{I}^2
- 2 \frac{Q}{I}\sigma_{QI}
- 2 \frac{U}{I}\sigma_{UI}
\end{aligned}}
\end{equation} where $\sigma_{Q}$ and $\sigma_{U}$ are the uncertainties in Stokes Q and U and where $\sigma_{QU}$, $\sigma_{QI}$, and $\sigma_{UI}$ are the covariance uncertainty terms. The polarization fraction is debiased ($p^\prime$), following the HAWC+ handbook \citep{hawcplus_handbook} by $p^\prime = \sqrt{p^2 - \sigma_{p}^2}$.
From all of these, the debiased polarization intensity ($P^\prime = I \times p^\prime/100$) and its errors are calculated. 
Although the uncertainties are calculated per pixel, the median Stokes I RMS is $\sigma_I$ = 2.2 mJy/arsec$^2$ and the median polarization intensity RMS $\sigma_p$ = 0.22 mJy/arsec$^2$.

For the final maps, we selected the debiased polarization vectors that met the criteria of $P^\prime$/$\sigma_{p}$ $>$ 3, the debiased polarization percentage was $<$ 50\%, and the Stokes I flux values were at least 
$\times$10 the Stokes I RMS noise level. While this was done at every pixel in the image, we present only enough vectors to be considered Nyquist sampled, for a total of 45 polarization vectors.

We estimate the overall flux calibration uncertainty of the observations at 20\%, but for the remainder of the paper, any flux uncertainty listed will only consider statistical uncertainty.  

\subsection{Archival Data}

To compare our polarization observations with other wavelengths, we used JCMT POL-2 850 $\mu$m wavelength polarization observations at 14$\arcsec$ resolution from \cite{pattle2021}, and ALMA Band 6 (1.3 mm wavelength) polarization observations from \citep{Sadavoy2018}, smoothed to 1$\arcsec$ resolution.

In addition, we also used Herschel 70, 100, and 160 $\mu$m PACS data\footnote{PACS observing labels 1342205093, 1342205094, 1342227150, 1342227151} and 250, 350, and 500 $\mu$m SPIRE data\footnote{SPIRE observing labels 	
1342205093, 	
1342205094} (Spectral and Photometric Imaging Receiver).

We included the 24 $\mu$m Spitzer MIPS data\footnote{MIPS observing label 4321536} (Multiband Imaging Photometer). All of these data were obtained from the Herschel Science Archive\footnote{http://archive.esac.esa.int/hsa/whsa/}. Lastly, the 22 $\mu$m WISE data\footnote{WISE observing coadd id: 2477m243\_ac51} (Wide-field Infrared Survey Explorer) from the NASA/IPAC Infrared Science Archive\footnote{https://irsa.ipac.caltech.edu0/applications/wise/} were also included.

\section{Results} \label{sec:results}

\input{90_polmap.tex}

Figure \ref{fig:90} shows the HAWC+ Band C (89 $\mu$m) dust continuum observations of IRAS 16293 with the   
inferred magnetic field polarization direction (i.e. polarization rotated by 90 degrees) plotted over the 89 $\mu$m continuum map as line segments, hereafter called vectors. 

We fit the protostellar envelope continuum emission using a Gaussian. The protostellar envelope is resolved with a deconvolved fit size of 10.3$^{\prime\prime}$ $\pm$ 0.4$^{\prime\prime}$ by 8.7$^{\prime\prime}$ $\pm$ 0.4$^{\prime\prime}$ with a position angle (PA) of 108$^{\circ}$ $\pm$ 11$^{\circ}$.  The integrated fit flux is 1236 $\pm$ 25 Jy, and the peak flux is 1.25 $\pm$ 0.02 Jy/arcsec$^{2}$.

The continuum Gaussian fit of the HAWC+ 89 $\mu$m observation is different than Gaussian fits to the Herschel archival data. The PACS 100 $\mu$m source fits are somewhat smaller  (7.7$^{\prime\prime}$ $\pm$ 0.4$^{\prime\prime}$ by 4.8$^{\prime\prime}$ $\pm$ 0.4$^{\prime\prime}$ and a PA of 141$^{\circ}$ $\pm$5$^{\circ}$), whereas the PACS 70 $\mu$m source fits are a little larger (11.9$^{\prime\prime}$ $\pm$ 0.3$^{\prime\prime}$ by 10.1$^{\prime\prime}$ $\pm$ 0.3$^{\prime\prime}$ and a PA of 113$^{\circ}$ $\pm$ 7$^{\circ}$). The variation indicated in the Gaussian fits of the HAWC+ and 70 and 100 PACS $\mu$m images likely arises from differences in the source morphologies (e.g., outflows, envelopes, and the bridge connecting source A and B) so close to the overall emission peak near 100 $\mu$m. On the other hand, the PA from the HAWC+ Band C fit is within 10$^{\circ}$ of the 850 $\mu$m dust emission \citep{pattle2021}, although the 850 $\mu$m core extension is measured on much larger scales.

\input{91_polmap_w_kates_lines.tex}

We can compare the inferred magnetic field from other observations with varying wavelength and spatial scales. Figure \ref{fig:91} shows the inferred magnetic field from our 89 $\mu$m data from Figure \ref{fig:90} (black vectors) with the 850 $\mu$m data from POL-2 (orange vectors), which have 14$\arcsec$ resolution and 12$\arcsec$ pixels \citep{pattle2021}, both wavelengths presented as normalized vectors.
The vectors align on the western side of the source (and one vector in the southeast), but otherwise there is generally not much agreement on the inferred magnetic field direction. 

On the other hand, misalignment is also seen when comparing the core scale (POL-2) with the cloud scale (Planck). 
As pointed out by \cite{pattle2021}, there is a misalignment of the large scale Planck field with the POL-2 850 $\mu$m magnetic field near the IRAS 16293 protostars, which is also seen in Figure \ref{fig:91} as a field shift near the protostars compared to the outer region.  
In Figure \ref{fig:90}, the inferred HAWC+ magnetic field is generally aligned EW.  This is offset by $\sim$65$^{\circ}$ from the overall $\sim$arcminute large-scale field inferred from Planck observations of the Ophiuchus L1689 molecular cloud core, which is 24$^{\circ}$  E of N \citep{pattle2021}. 

At higher spatial scale, Figure \ref{fig:93} 
compares the 89 $\mu$m HAWC+ data and 850 $\mu$m POL-2 data with high resolution polarization observations at 1.3 mm with ALMA from \cite{Sadavoy2018}, which are consistent with 880$\mu$m polarization observations from \cite{Rao2009}.  The ALMA data have been smoothed to 1$\arcsec$ resolution to better compare with our observations.

Note that the region mapped in  Figure \ref{fig:93} is only slightly larger than the POL-2 and HAWC+ beam sizes, so it is difficult to do more than compare broadly.
However, in general, the vectors from the three observations are not well-aligned. The HAWC+ and ALMA observations have some agreement to the northwest, and the ALMA and POL-2 observations have some agreement in the bridge region between source A and B. On the other hand, HAWC+ and ALMA observations are anti-aligned in the South, while ALMA and POL-2 observations are anti-aligned in the middle-eastern side.

\input{93_polmap_w_sarahs_lines.tex}

\section{Discussion} \label{sec:discussion}

\input{92_peak_map.tex}

\subsection{Continuum Wavelength Dependence} \label{4.1}

When comparing the HAWC+ dust continuum emission Gaussian fits with the Gaussian fits from Herschel, as well as the peak in the POL-2 observations, we noted that the peak of the emission varied with wavelength-- with the far infrared wavelengths peaking between the protostellar source locations. 
As expected, the flux peak dust continuum of the IRAS 16293-2422 system is strongly dependent on observing wavelength, as can be seen in the SED from \cite{schoier2002} with a maximum around 100 $\mu$m, but it also seems that the spatial location of the peak is very strongly dependent on observing wavelength.
By using the archival observations, we can compare this trend more broadly.  

We fit the different wavelength emission (e.g., WISE, Herschel, and SCUBA2 listed in Section \ref{sec:observations}) to Gaussians. Figure \ref{fig:92} shows the Gaussian fit peak to wavelengths from 22 $\mu$m to 850 $\mu$m. There is a clear gradient from northwest to southeast from short wavelengths to long wavelengths. In other words, we detect a gradient in the cloud dust temperatures with wavelength.  This is likely due to  differences in the morphology or detailed properties of the source, possibly arising from the structure of the envelope or disks in the binary system.

Another explanation of the shift with wavelength could be a difference in evolutionary stage of the binary sources. As stated in \S 1, source B has been argued to be more evolved than source A \citep{Stark2004}. As we probe shorter and shorter wavelengths, we are most sensitive to the hot dust in the less obscured source. 
However, the evolutionary stages of the protostars are still uncertain with most recent work suggesting that the sources are likely at about the same evolutionary stage or that source B is slightly less evolved \citep{vanderwiel2019}, but the evolutionary ages are still a point of contention. 

One of the complications is that the two protostars have different inclinations with respect to the line of sight. Source A, which is the  tight binary, is closer to edge-on \citep[inclinations of 59$^\circ$ and 74$^\circ$;][]{maureira2020} while source B is close to face-on \citep[inclination $\sim$18$^\circ$;][]{zamponi2021}.  The source A protostars are most likely Class 0 based on evidence of them driving  active outflows \citep[e.g.,][]{Yeh2008}, but the evolution of source B is harder to verify due to its face-on inclination and lack of outflow activity \citep[cf.,][]{Yeh2008}. The difference in inclination could explain the observed peak shift seen at the shorter wavelengths, since the shorter wavelengths are most sensitive to the hotter dust seen in the more face-on disk, which is less obscured. In addition, source A is also surrounded by a $\sim$100 au circumbinary disk with a PA of $\sim$50$^\circ$ \citep{maureira2020}, which could increase the  obscuration for the binary. We suggest that such continuum peak positional shifts with wavelength may be present in other binary sources, requiring careful consideration of variation in peak position at the shorter wavelengths.  

\input{94_polmap_w_yeh_contours.tex}

\subsection{Magnetic Fields} \label{4.2}

At the large scales in dense, elongated structures or filaments of molecular clouds, the magnetic field's alignment depends on column density: below a critical column density, the magnetic field is preferentially oriented parallel to the elongation of density structures, and above, the magnetic field is preferentially perpendicular \citep{planckXXXV2016}. However, the details of the alignment may vary across star formation regions \citep{Stephens2022}. Using HAWC+ observations in L1688, it has been shown that the transition occurs at a molecular hydrogen column density of $\sim$10$^{21.7}$ cm$^{-2}$ \citep{lee2021}.  Magneto-hydrodynamic simulations have suggested that the transition may coincide with the kinetic energy of the gravitationally induced flows surpassing the magnetic energy \citep{chen2016}.

One of the goals of our observations was to compare the inferred magnetic field in the far infrared dust continuum to other resolutions and scales that impact the star formation process.  The large scale Planck inferred magnetic field in the L1689-N region is 24$^{\circ}$ E of N \citep{pattle2021}, i.e. nearly perpendicular to the large 
L1689/L1712 filament identified in \cite{Ladjelate2020}.  These large-scale Planck fields are generally consistent with the 850 $\mu$m JCMT POL-2 inferred magnetic fields shown in Figure \ref{fig:91}, although in detail the POL-2 fields are less aligned with the overall Planck field at the cloud core \citep[e.g., see Figure 4 in][]{pattle2021}. This is not surprising as the overall Planck field seems to exhibit a small shift at the the core that suggests unresolved morphology. 

When one traces the magnetic field from the L1689-N core scale down to the protostellar envelope scale with the three observations in Figure \ref{fig:93}, we see regions where the inferred magnetic fields are aligned but other regions where the inferred magnetic fields  are misaligned.
The HAWC+ and ALMA observations are aligned in the northwest, while the ALMA and POL-2 observations are well aligned in the bridge region between source A and B. On the other hand, HAWC+ and ALMA observations are misaligned in the South, while ALMA and POL-2 observations are misaligned on the eastern side.
However, as pointed out by \cite{pattle2021}, the average field angles are consistent between ALMA and POL-2: 166$^\circ\pm$31$^\circ$ for POL-2 at the cloud core and 176$^\circ\pm$54$^\circ$ and 130$^\circ\pm$14$^\circ$ averaged around the protostars only and the bridge between them only in the unsmoothed ALMA data in \cite{Sadavoy2018}, respectively. 
Whereas the HAWC+ observations have a very different measured polarization angle: 89$^\circ\pm$23$^\circ$, which is more in line with the E-W outflow.

Indeed, one of the most striking features of the HAWC+ map is the uniform field that is nearly completely EW in direction, which is very different morphologically to the Planck, JCMT, and ALMA inferred magnetic field angles.  Although one could argue for large field morphological connections  or trends between the mapped fields in Figure \ref{fig:93}, there is still not overall coherence across wavelengths. The key to understanding these different wavelength observations is that the observations are probing different optical depth and dust temperatures (as seen clearly in Figure \ref{fig:93}), making it difficult to piece together the overall 3D magnetic field morphology without a better understanding of the system details. 

The IRAS 16293-2422 triple system has a complicated morphology with the protostars, disks, a bridge feature, and multiple outflows observed on different spatial scales \citep[see Figure 1 in][]{vanderwiel2019}. To date, the outflows have only been associated with source A, which is likely due to the face-on inclination of source B or possibly a difference in evolutionary stage. 
As discussed in \S 1, source A has two outflows on the 1000s of au scale: one east-west \citep[e.g.,][]{Yeh2008} and one northwest-southeast \citep[e.g.,][]{kristensen2013}. Arguably, the east-west outflow dominates the system on the cloud core scale \citep[e.g.,][]{mizuno1990,Stark2004}, which is also well detected in higher resolution interferometric observations \citep{Yeh2008}. Figure \ref{fig:94} shows our inferred magnetic fields with the blue- and red-shifted outflows as identified from CO (3-2) observations of \cite{Yeh2008}. Our magnetic field vectors are well aligned with this outflow direction, with the exception of our vectors extending more in the south. Nonetheless, based on this comparison, we posit that the HAWC+ observations are tracing the warm dust from the E-W outflow cavity or cavity walls of IRAS 16293-2422 \citep{Stark2004}.
This is in contrast to the average magnetic field from the POL-2 and ALMA observations, which are approximately perpendicular to the E-W outflow.  On the other hand, this is consistent with the \cite{Huang2024} survey, where $\sim$40\% of the sources have average magnetic fields perpendicular to the source outflow.

There are many morphological examples of the outflows of protostars seeming to modify or shape the magnetic field of the core as traced by single dish observations: Orion A filament \citep{Pattle2017}, NGC 2071IR in Orion B \citep{Lyo2021}, and 
CB 54 \citep{Pattle2022}
or to modify or shape the inner protostellar envelope as traced by interferometric observations:
Serpens SMM1 \citep{Hull2017}, B335 \citep{Maury2018}, Emb 8(N) \citep{LeGouellec2019}, and BHR 71 IRS2 \citep{Hull2020}. 

Additionally, the short wavelength observational inferred magnetic field can exhibit different morphologies compared to the longer wavelengths. In Orion, for example, the HAWC+ shorter wavelengths (53 and 89 $\mu$m compared to 154 and 214 $\mu$m) is more aligned with the bipolar outflow structure in the Becklin–Neugebauer/Kleinman–Low region, as traced by molecular tracers \citep{chuss2019}. The longer wavelengths are argued to be tracing the cooler dust that is outside of the explosion influenced region.

In this case for IRAS 16293-2422, the 89 $\mu$m continuum emission is tracing the warmer dust surrounding the outflow. The magnetic field morphology in this region is dominated by the outflow and not the magnetic field in the cloud core or the bridge, which is seen to dominate at longer wavelengths.
This is somewhat consistent with the continuum peak of IRAS 16293-2442 at $\sim$100 $\mu$m being near the center of the system so that the polarization is dominated by the large opening angle cavity seen in Figure \ref{fig:94}.

\subsection{Possibility of Polarization by Dichroic Extinction}

As the inferred magnetic direction in Figure \ref{fig:90} is different than what is observed at 850 $\mu$m with POL-2, we should also examine other polarization mechanisms that may account for the observations.  The most likely other polarization mechanism would be dichroic extinction, used in the optical or infrared to infer magnetic fields from polarization observations of stars extincted by aligned dust grains \citep[e.g.,][]{scarrott1989}.  Although dichroic extinction depends on dust properties, optical depth, and temperature gradients \citep[also see,][]{hildebrand2000}, we can consider the effect in the typical case for young stellar objects when unpolarized, or weakly polarized, emission is extincted by cooler foreground dust that has its grains aligned by magnetic fields or other means. In that case, the dust grain's long axis will have more efficient extinction than the short axis, so the light is now polarized along the dust's short axis, which would result in a 90$^\circ$ flip in the polarization, compared to the inferred magnetic field from  emission of aligned dust grains. 
The effect has been seen in far infrared dust polarization observations, particularly in the Sagittarius B2 molecular cloud \citep{novak1997,dowell1997}.
The process will be most effective in optically thick sources with a temperature gradient. 

Although, we can not rule out dichroic extinction as the dominate mechanism for polarization in IRAS 16293-2422 
\citep[see][where they argue that dichroic extinction polarization is not important in HAWC+ observations of B335]{Zielinski2021}, there are a few reasons to disfavor that explanation. First, we do not see a strong signature of the 90$^\circ$ flip toward the outside of the source, as is seen, for example, in observations of NGC 1333 IRAS4 \citep{ko2020} and OMC-3/MMS 6 \citep{baobab2021}. In fact, with the exception of two vectors, Figure \ref{fig:90} shows a very uniform inferred magnetic field.  Of course such a flip is not necessary, but if we flip all of our vectors by 90$^\circ$, then the vectors on the west in Figure \ref{fig:91} that currently agree with the POL-2 and ALMA vectors would disagree. 
Second, although a flip of 90$^\circ$ for the inferred magnetic field would mean an average field of 179$^\circ\pm$23$^\circ$, which is closer to the bulk average magnetic fields of 166$^\circ\pm$31$^\circ$ for POL-2 and 176$^\circ\pm$54$^\circ$ for ALMA data, in detail the vectors in the inner region of the source, see Figure \ref{fig:93} for example, would not better match with a HAWC+ flip of 90$^\circ$. 
Third, the polarization fraction is lower toward the center of our observations in Figure \ref{fig:90}. Such polarization depression with increasing density are commonly detected in magnetic field aligned dust grain polarization observations \citep[e.g.,][]{chuss2019} and are not seen in regions where the polarization is due to dichroic extinction \citep{novak1997,dowell1997,ko2020}.
Finally, east-west magnetic fields have also been detected along a few stellar sight lines in the L1689 cloud at optical wavelengths \citep[e.g.,][]{Vrba1976}, suggesting that there are east-west fields in some parts of the region. 
Overall, based on these four reasons, the IRAS 16293-2422 polarization observations presented here are most likely due to magnetic field alignment of dust grains.

\section{Conclusions}

We present the 89 $\mu$m continuum polarization emission toward the protostellar system IRAS 16293-2422 using the HAWC+ polarimeter onboard SOFIA. Our main conclusions are:

\begin{enumerate}

\item We detect a uniform magnetic field in the inner region of IRAS 16293-2422 that is aligned east-west.  The average field angle in the HAWC+ observations is 89$^\circ\pm$23$^\circ$, which is consistent with one of the known large scale outflows.
This is different from the average field angles at longer wavelengths: Planck large-scale average field angle of 24$^\circ$, JCMT POL-2 average field angle
of 166$^\circ\pm$31$^\circ$, and the ALMA average field angle near the protostars of 176$^\circ\pm$54$^\circ$.

\item We posit that the magnetic field probed by the 89 $\mu$m continuum emission is dominated by the outflow magnetic field, while the 850 $\mu$m dust emission is dominated more by the core magnetic field.

\item The continuum peak of the source varies significantly with wavelength, moving from near Source B at the midIR to near Source A in the FIR. This is either a consequence of evolutionary state of the two sources or more likely due to the difference in inclination. The shorter wavelengths are more sensitive to the warmer dust which is seen in Source B.

\end{enumerate}

These observations suggest that magnetic fields in young protostars are likely more complicated than simple pictures suggest with multiple field morphologies dominating at various scales and in different structures.  Any observational constraints should include multiwavelength observations that sample multiple scales and all the structures of star formation.

\acknowledgments
Based on observations made with the NASA/DLR Stratospheric Observatory for Infrared Astronomy (SOFIA). SOFIA is jointly operated by the Universities Space Research Association, Inc. (USRA), under NASA contract NNA17BF53C, and the Deutsches SOFIA Institut (DSI) under DLR contract 50 OK 0901 to the University of Stuttgart.
L.W.L. acknowledges support from NSF AST-1910364 and NSF AST-2307844. R.E.H acknowledges support from NSF AST-1910364. 
G.N. is grateful for NASA support for this work, via award Nos. SOF 06-0116 and SOF 07-0147 issued by USRA to Northwestern University. 
E.G.C. acknowledges support from the National Science Foundation through the NSF MPS Ascend Fellowship Grant number 2213275. 
K.P. is a Royal Society University Research Fellow, supported by grant no. URF\textbackslash R1\textbackslash211322.
We thank Naomi Hirano for providing us with the calibrated SMA CO observations in Figure \ref{fig:94}. We thank John Tobin for discussions on binarity and Zhi-Yun Li for discussions on magnetic field alignment with column density. 

\bibliography{8_bib}{}
\bibliographystyle{aasjournal}

\end{document}

%% file: 90_polmap.tex
\begin{figure}[]
  \centering
    \includegraphics[width=0.45\textwidth,trim={2 2 2 2},clip]{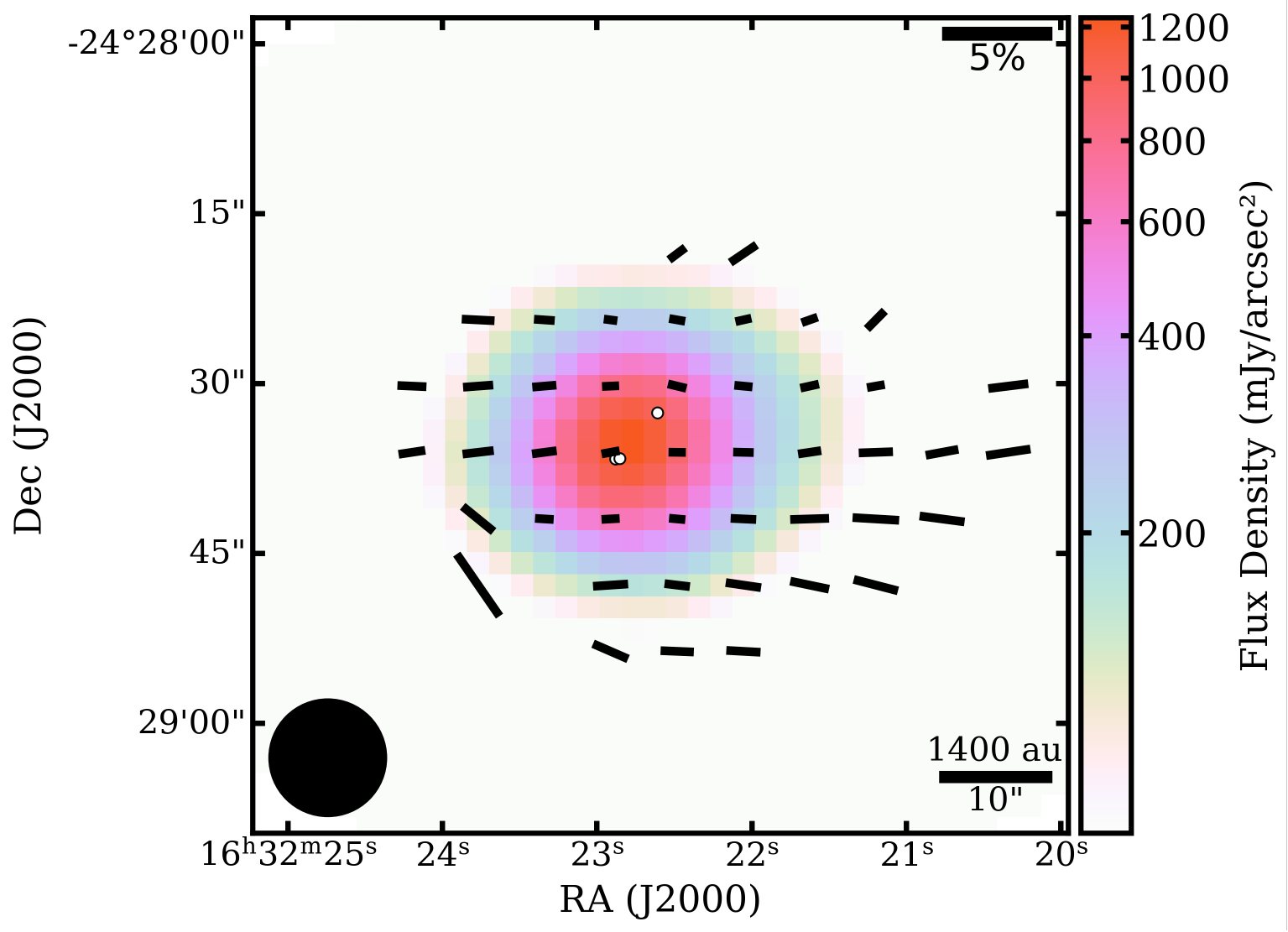}
  \caption{HAWC+ 89 $\mu$m continuum map overlaid with the inferred magnetic field direction in black. The three protostellar peaks at 3 mm are shown in white \cite[]{maureira2020}, source A1/A2 in the South and source B in the North. The vectors were selected to show Nyquist sampling. The image has a smoothed image beam size of 11.7$^{\prime\prime}$. The polarization scale bar is in the top right corner, the beam is in the bottom left corner, and a spatial scalebar of 10$^{\prime\prime}$ (1400 au) is provided in the bottom right corner.}
  \label{fig:90}
\end{figure}

%% file: 91_polmap_w_kates_lines.tex
\begin{figure}[]
  \centering
    \includegraphics[width=0.45\textwidth,trim={2 3 2 2},clip]{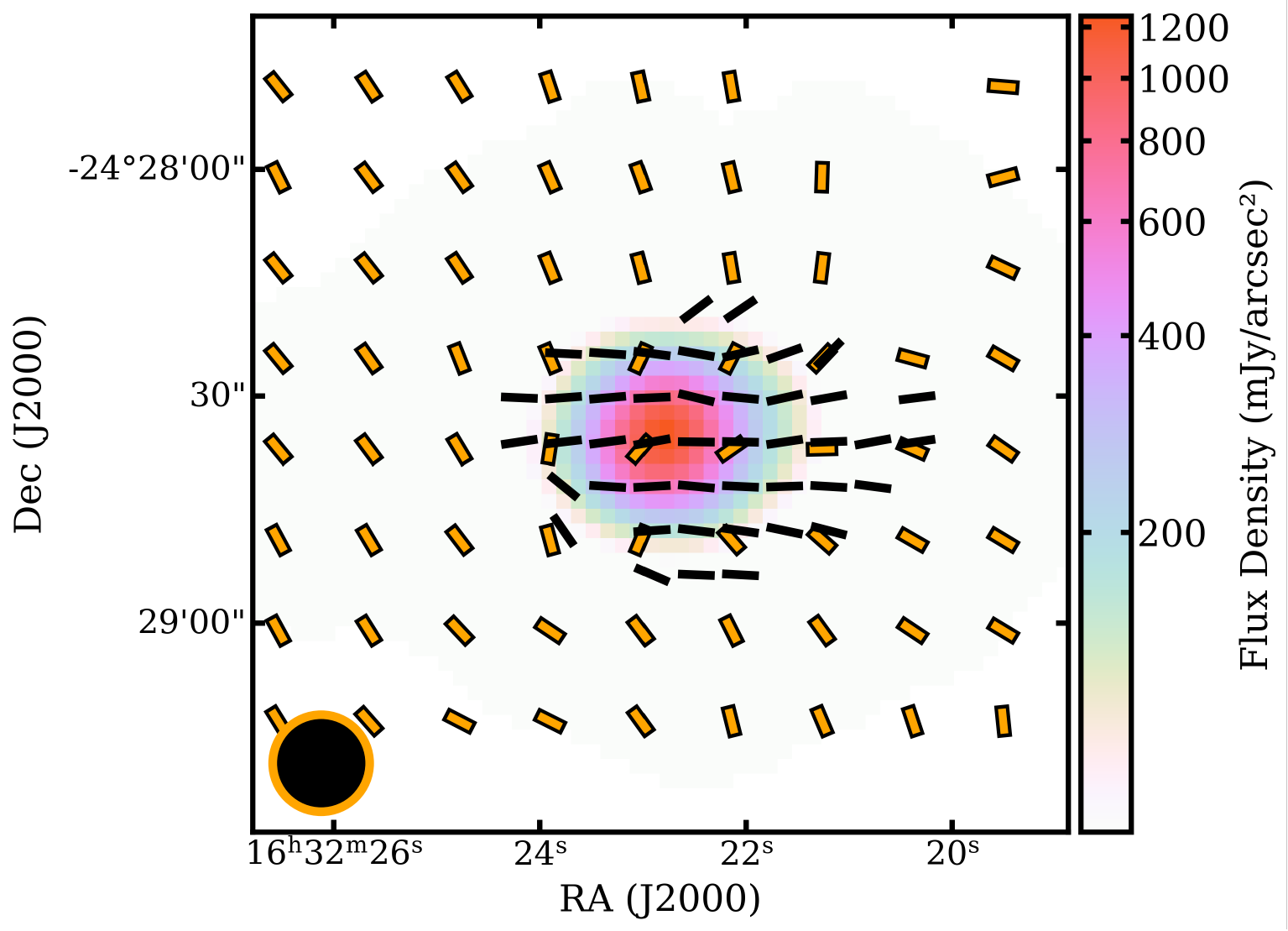}
  \caption{Our continuum map overlaid with our normalized, Nyquist sampled inferred magnetic field vectors in black. In orange vectors, we show the 850 $\mu$m POL-2 normalized-vectors from \cite{pattle2021} at the published pixel scale of 12$^{\prime\prime}$. The corresponding beams, 11.7$\arcsec$ and 14$\arcsec$ for our observation and POL-2, respectively, are in the bottom left corner.}
  \label{fig:91}
\end{figure}

%% file: 93_polmap_w_sarahs_lines.tex
\begin{figure}[]
  \centering
    \includegraphics[width=0.45\textwidth,trim={2 3 2 2},clip]{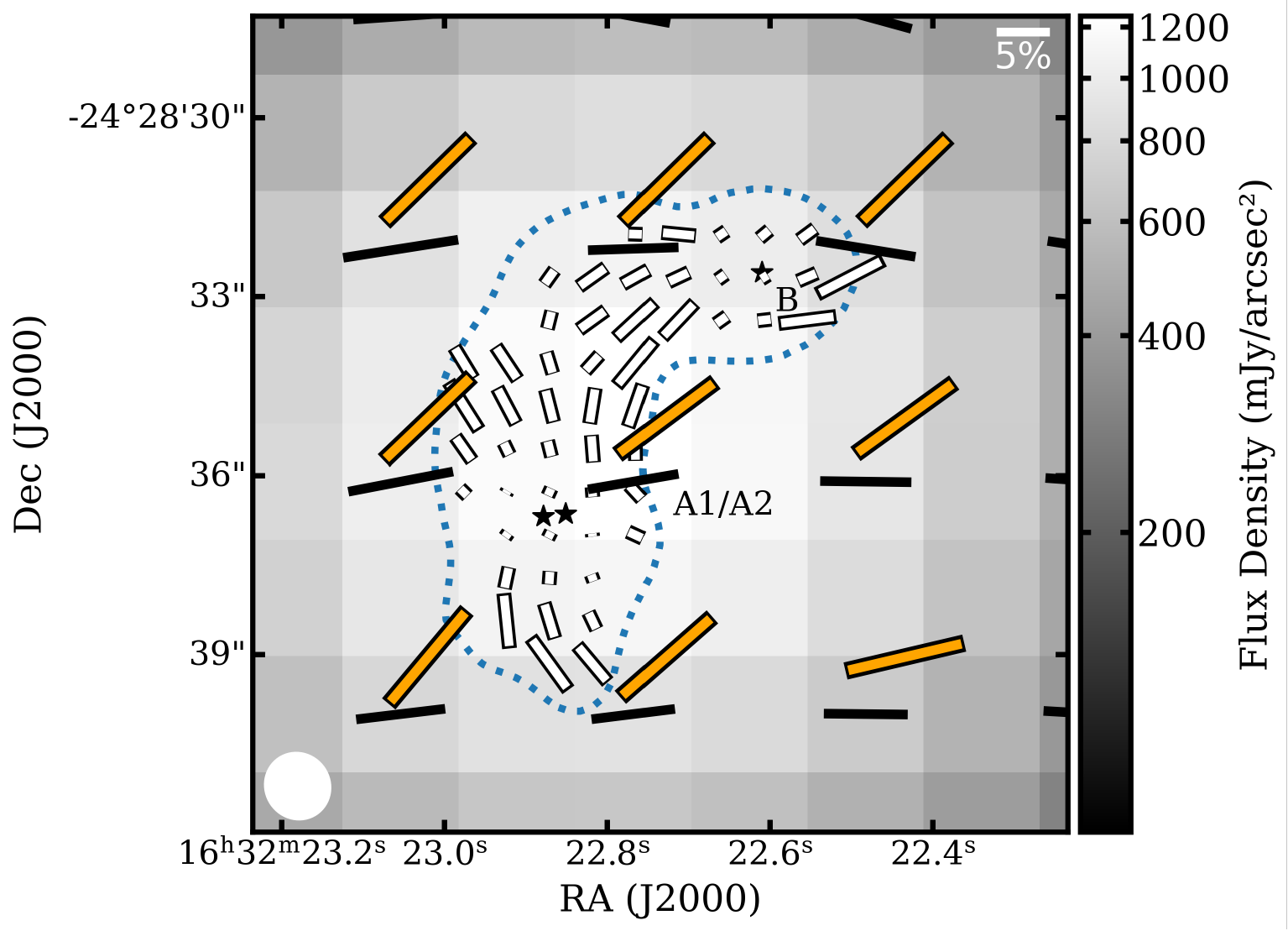}
  \caption{
  The HAWC+ 89 $\mu$m continuum image overlaid with the normalized magnetic field vectors from HAWC+ in black and from POL-2 in orange.  Both maps show oversampled vectors at 4$\arcsec$ pixels to better match the ALMA scales.  The white vectors show the ALMA 1.3 mm polarization vectors, smoothed to 1$\arcsec$ resolution (white beam in the bottom left corner).  The ALMA vectors are not normalized, and a 5\% scale bar is given in the top right corner.  The dotted contour shows the ALMA continuum at the 10$\sigma_I$ level, where $\sigma_I$ = 1.72 mJy arcsec$^{-1}$. The locations of the hierarchical triple protostars (source A1/A2 and source B) at 3 mm \cite[e.g.,][]{maureira2020} are labeled. The bridge is the region connecting the two sources}
  \label{fig:93}
\end{figure}

%% file: 92_peak_map.tex
\begin{figure}[]
  \centering
    \includegraphics[width=0.45\textwidth,trim={2 3 2 2},clip]{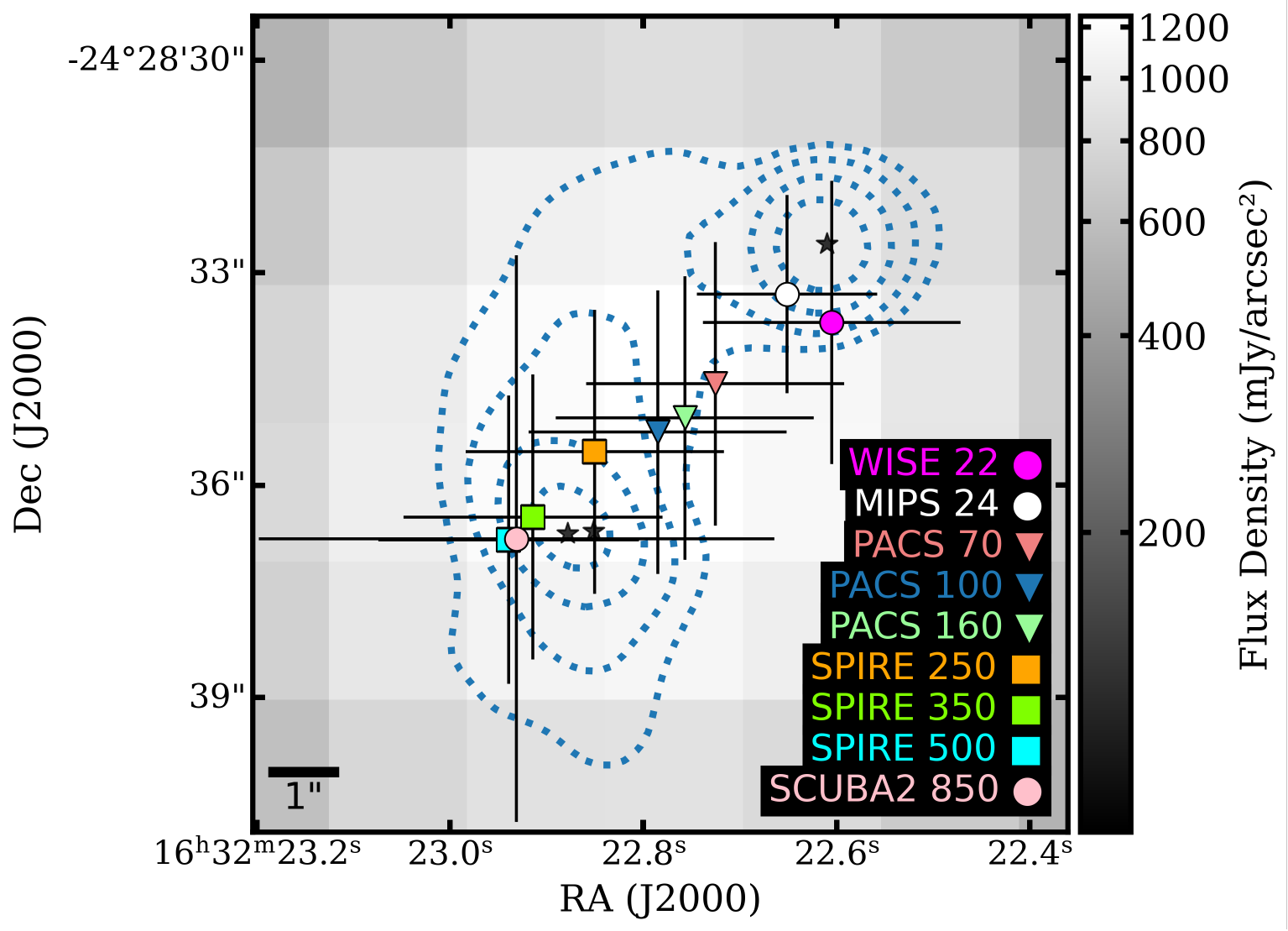}
  \caption{Image of the IRAS 16293-2422 system with Gaussian-fit peak positions observed by various telescopes/instruments. Background image is our HAWC+ Stokes I map. From right to left: the magenta marker is the Gaussian peak of the {\it WISE} W4 filter at 22 $\mu$m; the white marker is {\it Spitzer} MIPS 24 $\mu$m peak; light coral, pale green, and blue corresponding to {\it Herschel} PACS at 70, 160, and 100 $\mu$m; orange and chartreuse for SPIRE at 250 and 350 $\mu$m; the pink marker corresponding to SCUBA-2  at 850 $\mu$m; and the cyan is SPIRE at 500 $\mu$m. Errors represent pointing uncertainties collected from the respective telescope's manual, main publications, or relevant sources (WISE \citep{wright2010}, MIPS on {\it Spitzer} \citep{mips_handbook}, PACS \& SPIRE on {\it Herschel} \citep{sanchez_portal2014}, SCUBA2 on JCMT \citep{Pattle2015,helen2018}). Contours are 1.3 mm continuum from ALMA (smoothed to 1$^{\prime\prime}$) with levels at N$\sigma$ for N$\in$[10, 30, 100, 300] and $\sigma$ = 1.72 mJy arcsec$^{-1}$ \citep{Sadavoy2018}.}
  \label{fig:92}
\end{figure}

%% file: 94_polmap_w_yeh_contours.tex
\begin{figure}[]
  \centering
    \includegraphics[width=0.45\textwidth,trim={2 3 2 2},clip]{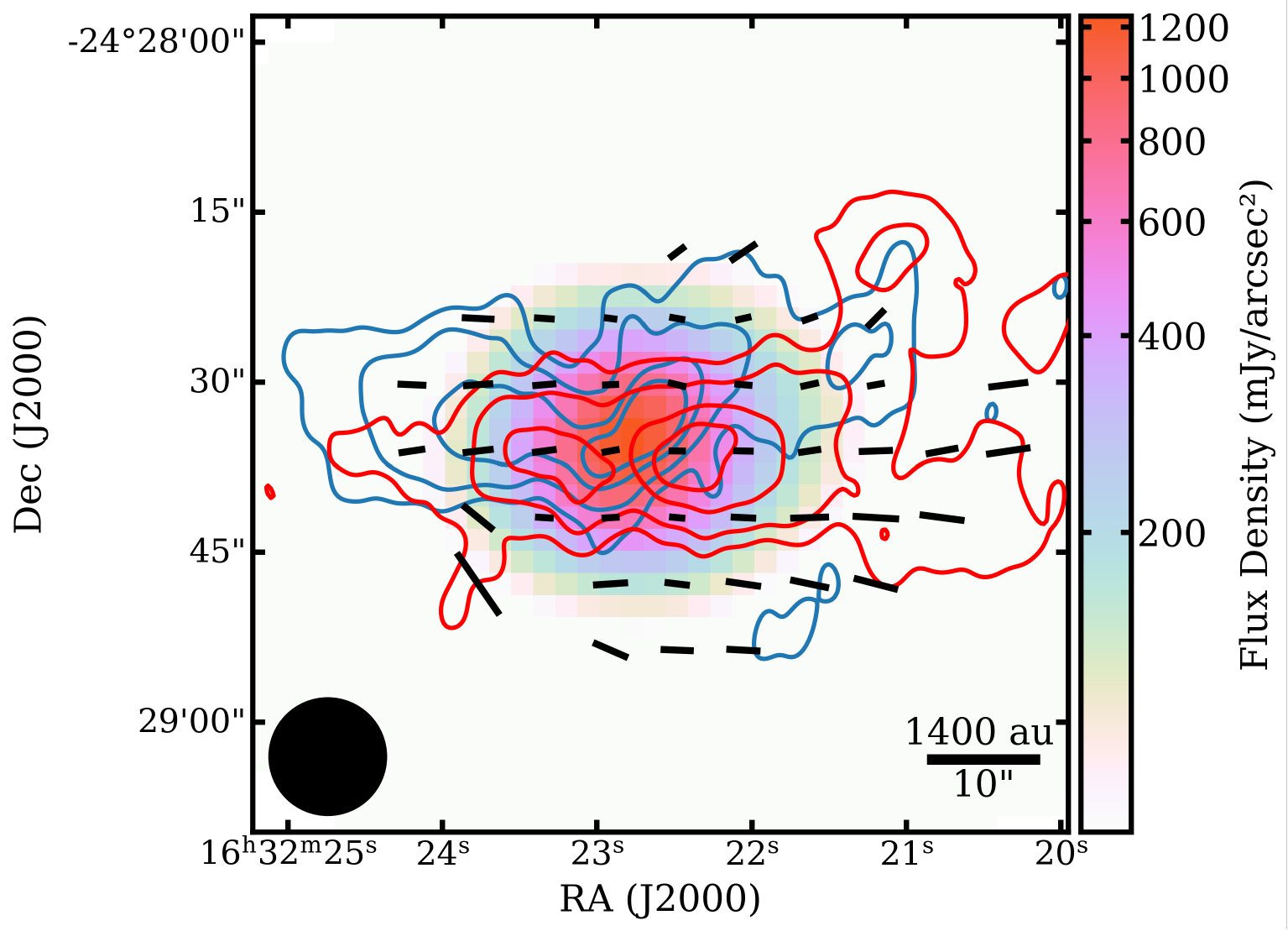}
  \caption{Same as Figure \ref{fig:90} but overlaid with the blue and red $^{12}$CO 3-2 outflow contour lines of \cite{Yeh2008} using contours at 3, 10, 30, and 50$\sigma$, where $\sigma$ is 1.4 Jy beam$^{-1}$ for the blue and 2.3 Jy beam$^{-1}$ for the red.}
  \label{fig:94}
\end{figure}

%% file: main.bbl
\begin{thebibliography}{}
\expandafter\ifx\csname natexlab\endcsname\relax\def\natexlab#1{#1}\fi
\providecommand{\url}[1]{\href{#1}{#1}}
\providecommand{\dodoi}[1]{doi:~\href{http://doi.org/#1}{\nolinkurl{#1}}}
\providecommand{\doeprint}[1]{\href{http://ascl.net/#1}{\nolinkurl{http://ascl.net/#1}}}
\providecommand{\doarXiv}[1]{\href{https://arxiv.org/abs/#1}{\nolinkurl{https://arxiv.org/abs/#1}}}

\bibitem[{{Adams} {et~al.}(1989){Adams}, {Ruden}, \& {Shu}}]{adams1989}
{Adams}, F.~C., {Ruden}, S.~P., \& {Shu}, F.~H. 1989, \apj, 347, 959, \dodoi{10.1086/168187}

\bibitem[{{Chapman} {et~al.}(2013){Chapman}, {Davidson}, {Goldsmith}, {Houde}, {Kwon}, {Li}, {Looney}, {Matthews}, {Matthews}, {Novak}, {Peng}, {Vaillancourt}, \& {Volgenau}}]{Chapman2013}
{Chapman}, N.~L., {Davidson}, J.~A., {Goldsmith}, P.~F., {et~al.} 2013, \apj, 770, 151, \dodoi{10.1088/0004-637X/770/2/151}

\bibitem[{{Chen} {et~al.}(2016){Chen}, {King}, \& {Li}}]{chen2016}
{Chen}, C.-Y., {King}, P.~K., \& {Li}, Z.-Y. 2016, \apj, 829, 84, \dodoi{10.3847/0004-637X/829/2/84}

\bibitem[{{Chuss} {et~al.}(2019){Chuss}, {Andersson}, {Bally}, {Dotson}, {Dowell}, {Guerra}, {Harper}, {Houde}, {Jones}, {Lazarian}, {Lopez Rodriguez}, {Michail}, {Morris}, {Novak}, {Siah}, {Staguhn}, {Vaillancourt}, {Volpert}, {Werner}, {Wollack}, {Benford}, {Berthoud}, {Cox}, {Crutcher}, {Dale}, {Fissel}, {Goldsmith}, {Hamilton}, {Hanany}, {Henning}, {Looney}, {Moseley}, {Santos}, {Stephens}, {Tassis}, {Trinh}, {Van Camp}, {Ward-Thompson}, \& {HAWC + Science Team}}]{chuss2019}
{Chuss}, D.~T., {Andersson}, B.~G., {Bally}, J., {et~al.} 2019, \apj, 872, 187, \dodoi{10.3847/1538-4357/aafd37}

\bibitem[{Clarke {et~al.}(2020)Clarke, Berthoud, Kov{\'{a}}cs, Santos, \& Novak}]{hawcplus_handbook}
Clarke, M., Berthoud, M., Kov{\'{a}}cs, A., Santos, F., \& Novak, G. 2020, {Guest Observer Handbook for HAWC+ Data Products}, Rev. F.
\newblock \url{https://www.sofia.usra.edu/sites/default/files/Instruments/HAWC_PLUS/Documents/hawc_data_handbook.pdf}

\bibitem[{{Cox} {et~al.}(2018){Cox}, {Harris}, {Looney}, {Li}, {Yang}, {Tobin}, \& {Stephens}}]{cox2018}
{Cox}, E.~G., {Harris}, R.~J., {Looney}, L.~W., {et~al.} 2018, \apj, 855, 92, \dodoi{10.3847/1538-4357/aaacd2}

\bibitem[{{Cox} {et~al.}(2022){Cox}, {Novak}, {Sadavoy}, {Looney}, {Lee}, {Berthoud}, {Bourke}, {Coud{\'e}}, {Encalada}, {Fissel}, {Harrison}, {Houde}, {Li}, {Myers}, {Pattle}, {Santos}, {Stephens}, {Wang}, \& {Wolf}}]{Erin2022}
{Cox}, E.~G., {Novak}, G., {Sadavoy}, S.~I., {et~al.} 2022, \apj, 932, 34, \dodoi{10.3847/1538-4357/ac722a}

\bibitem[{{Crimier} {et~al.}(2010){Crimier}, {Ceccarelli}, {Maret}, {Bottinelli}, {Caux}, {Kahane}, {Lis}, \& {Olofsson}}]{crimier2010}
{Crimier}, N., {Ceccarelli}, C., {Maret}, S., {et~al.} 2010, \aap, 519, A65, \dodoi{10.1051/0004-6361/200913112}

\bibitem[{{Dowell}(1997)}]{dowell1997}
{Dowell}, C.~D. 1997, \apj, 487, 237, \dodoi{10.1086/304577}

\bibitem[{{Dzib} {et~al.}(2018){Dzib}, {Ortiz-Le{\'o}n}, {Hern{\'a}ndez-G{\'o}mez}, {Loinard}, {Mioduszewski}, {Claussen}, {Menten}, {Caux}, \& {Sanna}}]{dzib2018}
{Dzib}, S.~A., {Ortiz-Le{\'o}n}, G.~N., {Hern{\'a}ndez-G{\'o}mez}, A., {et~al.} 2018, \aap, 614, A20, \dodoi{10.1051/0004-6361/201732093}

\bibitem[{{Encalada} {et~al.}(2021){Encalada}, {Looney}, {Tobin}, {Sadavoy}, {Segura-Cox}, {Cox}, {Li}, \& {Novak}}]{encalada2021}
{Encalada}, F.~J., {Looney}, L.~W., {Tobin}, J.~J., {et~al.} 2021, \apj, 913, 149, \dodoi{10.3847/1538-4357/abf4fd}

\bibitem[{{Fukui} {et~al.}(1986){Fukui}, {Sugitani}, {Takaba}, {Iwata}, {Mizuno}, {Ogawa}, \& {Kawabata}}]{fukui1986}
{Fukui}, Y., {Sugitani}, K., {Takaba}, H., {et~al.} 1986, \apjl, 311, L85, \dodoi{10.1086/184803}

\bibitem[{{Galametz} {et~al.}(2018){Galametz}, {Maury}, {Girart}, {Rao}, {Zhang}, {Gaudel}, {Valdivia}, {Keto}, \& {Lai}}]{Galametz2018}
{Galametz}, M., {Maury}, A., {Girart}, J.~M., {et~al.} 2018, \aap, 616, A139, \dodoi{10.1051/0004-6361/201833004}

\bibitem[{{Girart} {et~al.}(2014){Girart}, {Estalella}, {Palau}, {Torrelles}, \& {Rao}}]{Girart2014}
{Girart}, J.~M., {Estalella}, R., {Palau}, A., {Torrelles}, J.~M., \& {Rao}, R. 2014, \apjl, 780, L11, \dodoi{10.1088/2041-8205/780/1/L11}

\bibitem[{Harper {et~al.}(2018)Harper, Runyan, Dowell, Wirth, Amato, Ames, Amiri, Banks, Bartels, Benford, Berthoud, Buchanan, Casey, Chapman, Chuss, Cook, Derro, Dotson, Evans, Fixsen, Gatley, Guerra, Halpern, Hamilton, Hamlin, Hansen, Heimsath, Hermida, Hilton, Hirsch, Hollister, Hostetter, Irwin, Jhabvala, Jhabvala, Kastner, Kov{\'{a}}cs, Lin, Loewenstein, Looney, Lopez-Rodriguez, Maher, Michail, Miller, Moseley, Novak, Pernic, Rennick, Rhody, Sandberg, Sandford, Santos, Shafer, Sharp, Shirron, Siah, Silverberg, Sparr, Spotz, Staguhn, Toorian, Towey, Tuttle, Vaillancourt, Voellmer, Volpert, i~Wang, \& Wollack}]{harper2018}
Harper, D.~A., Runyan, M.~C., Dowell, C.~D., {et~al.} 2018, Journal of Astronomical Instrumentation, 07, 1840008, \dodoi{10.1142/s2251171718400081}

\bibitem[{{Harris} {et~al.}(2018){Harris}, {Cox}, {Looney}, {Li}, {Yang}, {Fern{\'a}ndez-L{\'o}pez}, {Kwon}, {Sadavoy}, {Segura-Cox}, {Stephens}, \& {Tobin}}]{harris2018}
{Harris}, R.~J., {Cox}, E.~G., {Looney}, L.~W., {et~al.} 2018, \apj, 861, 91, \dodoi{10.3847/1538-4357/aac6ec}

\bibitem[{{Hildebrand} {et~al.}(2000){Hildebrand}, {Davidson}, {Dotson}, {Dowell}, {Novak}, \& {Vaillancourt}}]{hildebrand2000}
{Hildebrand}, R.~H., {Davidson}, J.~A., {Dotson}, J.~L., {et~al.} 2000, \pasp, 112, 1215, \dodoi{10.1086/316613}

\bibitem[{{Houde} \& {Vaillancourt}(2007)}]{houde2007}
{Houde}, M., \& {Vaillancourt}, J.~E. 2007, \pasp, 119, 871, \dodoi{10.1086/521109}

\bibitem[{{Huang} {et~al.}(2024){Huang}, {Girart}, {Stephens}, {Fern{\'a}ndez L{\'o}pez}, {Arce}, {Carpenter}, {Cortes}, {Cox}, {Friesen}, {Le Gouellec}, {Hull}, {Karnath}, {Kwon}, {Li}, {Looney}, {Megeath}, {Myers}, {Murillo}, {Pineda}, {Sadavoy}, {S{\'a}nchez-Monge}, {Sanhueza}, {Tobin}, {Zhang}, {Jackson}, \& {Segura-Cox}}]{Huang2024}
{Huang}, B., {Girart}, J.~M., {Stephens}, I.~W., {et~al.} 2024, \apjl, 963, L31, \dodoi{10.3847/2041-8213/ad27d4}

\bibitem[{{Hull} {et~al.}(2020){Hull}, {Le Gouellec}, {Girart}, {Tobin}, \& {Bourke}}]{Hull2020}
{Hull}, C. L.~H., {Le Gouellec}, V. J.~M., {Girart}, J.~M., {Tobin}, J.~J., \& {Bourke}, T.~L. 2020, \apj, 892, 152, \dodoi{10.3847/1538-4357/ab5809}

\bibitem[{{Hull} {et~al.}(2014){Hull}, {Plambeck}, {Kwon}, {Bower}, {Carpenter}, {Crutcher}, {Fiege}, {Franzmann}, {Hakobian}, {Heiles}, {Houde}, {Hughes}, {Lamb}, {Looney}, {Marrone}, {Matthews}, {Pillai}, {Pound}, {Rahman}, {Sandell}, {Stephens}, {Tobin}, {Vaillancourt}, {Volgenau}, \& {Wright}}]{Hull2014}
{Hull}, C. L.~H., {Plambeck}, R.~L., {Kwon}, W., {et~al.} 2014, \apjs, 213, 13, \dodoi{10.1088/0067-0049/213/1/13}

\bibitem[{{Hull} {et~al.}(2017){Hull}, {Girart}, {Tychoniec}, {Rao}, {Cort{\'e}s}, {Pokhrel}, {Zhang}, {Houde}, {Dunham}, {Kristensen}, {Lai}, {Li}, \& {Plambeck}}]{Hull2017}
{Hull}, C. L.~H., {Girart}, J.~M., {Tychoniec}, {\L}., {et~al.} 2017, \apj, 847, 92, \dodoi{10.3847/1538-4357/aa7fe9}

\bibitem[{{Jacobsen} {et~al.}(2018){Jacobsen}, {J{\o}rgensen}, {van der Wiel}, {Calcutt}, {Bourke}, {Brinch}, {Coutens}, {Drozdovskaya}, {Kristensen}, {M{\"u}ller}, \& {Wampfler}}]{jacobsen2018}
{Jacobsen}, S.~K., {J{\o}rgensen}, J.~K., {van der Wiel}, M.~H.~D., {et~al.} 2018, \aap, 612, A72, \dodoi{10.1051/0004-6361/201731668}

\bibitem[{{J{\o}rgensen} {et~al.}(2016){J{\o}rgensen}, {van der Wiel}, {Coutens}, {Lykke}, {M{\"u}ller}, {van Dishoeck}, {Calcutt}, {Bjerkeli}, {Bourke}, {Drozdovskaya}, {Favre}, {Fayolle}, {Garrod}, {Jacobsen}, {{\"O}berg}, {Persson}, \& {Wampfler}}]{jorgensen2016}
{J{\o}rgensen}, J.~K., {van der Wiel}, M.~H.~D., {Coutens}, A., {et~al.} 2016, \aap, 595, A117, \dodoi{10.1051/0004-6361/201628648}

\bibitem[{{Kataoka} {et~al.}(2015){Kataoka}, {Muto}, {Momose}, {Tsukagoshi}, {Fukagawa}, {Shibai}, {Hanawa}, {Murakawa}, \& {Dullemond}}]{kataoka2015}
{Kataoka}, A., {Muto}, T., {Momose}, M., {et~al.} 2015, \apj, 809, 78, \dodoi{10.1088/0004-637X/809/1/78}

\bibitem[{{Kirk} {et~al.}(2018){Kirk}, {Hatchell}, {Johnstone}, {Berry}, {Jenness}, {Buckle}, {Mairs}, {Rosolowsky}, {Di Francesco}, {Sadavoy}, {Currie}, {Broekhoven-Fiene}, {Mottram}, {Pattle}, {Matthews}, {Knee}, {Moriarty-Schieven}, {Duarte-Cabral}, {Tisi}, \& {Ward-Thompson}}]{helen2018}
{Kirk}, H., {Hatchell}, J., {Johnstone}, D., {et~al.} 2018, \apjs, 238, 8, \dodoi{10.3847/1538-4365/aada7f}

\bibitem[{{Ko} {et~al.}(2020){Ko}, {Liu}, {Lai}, {Ching}, {Rao}, \& {Girart}}]{ko2020}
{Ko}, C.-L., {Liu}, H.~B., {Lai}, S.-P., {et~al.} 2020, \apj, 889, 172, \dodoi{10.3847/1538-4357/ab5e79}

\bibitem[{{Kristensen} {et~al.}(2013){Kristensen}, {Klaassen}, {Mottram}, {Schmalzl}, \& {Hogerheijde}}]{kristensen2013}
{Kristensen}, L.~E., {Klaassen}, P.~D., {Mottram}, J.~C., {Schmalzl}, M., \& {Hogerheijde}, M.~R. 2013, \aap, 549, L6, \dodoi{10.1051/0004-6361/201220668}

\bibitem[{{Ladjelate} {et~al.}(2020){Ladjelate}, {Andr{\'e}}, {K{\"o}nyves}, {Ward-Thompson}, {Men'shchikov}, {Bracco}, {Palmeirim}, {Roy}, {Shimajiri}, {Kirk}, {Arzoumanian}, {Benedettini}, {Di Francesco}, {Fiorellino}, {Schneider}, {Pezzuto}, {Motte}, \& {Herschel Gould Belt Survey Team}}]{Ladjelate2020}
{Ladjelate}, B., {Andr{\'e}}, P., {K{\"o}nyves}, V., {et~al.} 2020, \aap, 638, A74, \dodoi{10.1051/0004-6361/201936442}

\bibitem[{{Lazarian}(2007)}]{lazarian2007}
{Lazarian}, A. 2007, \jqsrt, 106, 225, \dodoi{10.1016/j.jqsrt.2007.01.038}

\bibitem[{{Le Gouellec} {et~al.}(2019){Le Gouellec}, {Hull}, {Maury}, {Girart}, {Tychoniec}, {Kristensen}, {Li}, {Louvet}, {Cortes}, \& {Rao}}]{LeGouellec2019}
{Le Gouellec}, V. J.~M., {Hull}, C. L.~H., {Maury}, A.~J., {et~al.} 2019, \apj, 885, 106, \dodoi{10.3847/1538-4357/ab43c2}

\bibitem[{{Lee} {et~al.}(2019){Lee}, {Offner}, {Kratter}, {Smullen}, \& {Li}}]{lee2019}
{Lee}, A.~T., {Offner}, S. S.~R., {Kratter}, K.~M., {Smullen}, R.~A., \& {Li}, P.~S. 2019, \apj, 887, 232, \dodoi{10.3847/1538-4357/ab584b}

\bibitem[{{Lee} {et~al.}(2021){Lee}, {Berthoud}, {Chen}, {Cox}, {Davidson}, {Encalada}, {Fissel}, {Harrison}, {Kwon}, {Li}, {Li}, {Looney}, {Novak}, {Sadavoy}, {Santos}, {Segura-Cox}, \& {Stephens}}]{lee2021}
{Lee}, D., {Berthoud}, M., {Chen}, C.-Y., {et~al.} 2021, \apj, 918, 39, \dodoi{10.3847/1538-4357/ac0cf2}

\bibitem[{{Liu}(2021)}]{baobab2021}
{Liu}, H.~B. 2021, \apj, 914, 25, \dodoi{10.3847/1538-4357/abf8b6}

\bibitem[{{Looney} {et~al.}(2000){Looney}, {Mundy}, \& {Welch}}]{looney2000}
{Looney}, L.~W., {Mundy}, L.~G., \& {Welch}, W.~J. 2000, \apj, 529, 477, \dodoi{10.1086/308239}

\bibitem[{{Lynds}(1962)}]{lynd1962}
{Lynds}, B.~T. 1962, \apjs, 7, 1, \dodoi{10.1086/190072}

\bibitem[{{Lyo} {et~al.}(2021){Lyo}, {Kim}, {Sadavoy}, {Johnstone}, {Berry}, {Pattle}, {Kwon}, {Bastien}, {Onaka}, {Di Francesco}, {Kang}, {Furuya}, {Hull}, {Tamura}, {Koch}, {Ward-Thompson}, {Hasegawa}, {Hoang}, {Arzoumanian}, {Won Lee}, {Lee}, {Byun}, {Kirchschlager}, {Doi}, {Kim}, {Hwang}, {Diep}, {Fanciullo}, {Lee}, {Park}, {Yoo}, {Chung}, {Whitworth}, {Mairs}, {Soam}, {Liu}, {Tang}, {Coud{\'e}}, {Andr{\'e}}, {Bourke}, {Vivien Chen}, {Chen}, {Ping Chen}, {Chen}, {Ching}, {Cho}, {Choi}, {Choi}, {Chrysostomou}, {Dai}, {Dowell}, {Duan}, {Duan}, {Eden}, {Eswaraiah}, {Eyres}, {Fiege}, {Fissel}, {Franzmann}, {Friberg}, {Friesen}, {Fuller}, {Gledhill}, {Graves}, {Greaves}, {Griffin}, {Gu}, {Han}, {Hatchell}, {Hayashi}, {Houde}, {Inoue}, {Inutsuka}, {Iwasaki}, {Jeong}, {Kang}, {Kataoka}, {Kawabata}, {Kemper}, {Kim}, {Kim}, {Kim}, {Kim}, {Kirk}, {Kobayashi}, {K{\"o}nyves}, {Kusune}, {Kwon}, {Lacaille}, {Lai}, {Law}, {Lee}, {Lee}, {Lee}, {Li}, {Li}, {Li}, {Liu}, {Liu}, {Liu}, {Lu}, {Matsumura}, {Matthews},
  {Moriarty-Schieven}, {Nagata}, {Nakamura}, {Nakanishi}, {Bich Ngoc}, {Ohashi}, {Parsons}, {Peretto}, {Priestley}, {Pyo}, {Qian}, {Qiu}, {Rao}, {Rawlings}, {Rawlings}, {Retter}, {Richer}, {Rigby}, {Saito}, {Savini}, {Scaife}, {Seta}, {Shimajiri}, {Shinnaga}, {Tahani}, {Tang}, {Tomisaka}, {Tram}, {Tsukamoto}, {Viti}, {Wang}, {Wang}, {Xie}, {Yen}, {Yuan}, {Yun}, {Zenko}, {Zhang}, {Zhang}, {Zhang}, {Zhou}, {Zhu}, {de Looze}, {Dowell}, {Falle}, {Robitaille}, \& {van Loo}}]{Lyo2021}
{Lyo}, A.~R., {Kim}, J., {Sadavoy}, S., {et~al.} 2021, \apj, 918, 85, \dodoi{10.3847/1538-4357/ac0ce9}

\bibitem[{{Maureira} {et~al.}(2020){Maureira}, {Pineda}, {Segura-Cox}, {Caselli}, {Testi}, {Lodato}, {Loinard}, \& {Hern{\'a}ndez-G{\'o}mez}}]{maureira2020}
{Maureira}, M.~J., {Pineda}, J.~E., {Segura-Cox}, D.~M., {et~al.} 2020, \apj, 897, 59, \dodoi{10.3847/1538-4357/ab960b}

\bibitem[{{Maury} {et~al.}(2018){Maury}, {Girart}, {Zhang}, {Hennebelle}, {Keto}, {Rao}, {Lai}, {Ohashi}, \& {Galametz}}]{Maury2018}
{Maury}, A.~J., {Girart}, J.~M., {Zhang}, Q., {et~al.} 2018, \mnras, 477, 2760, \dodoi{10.1093/mnras/sty574}

\bibitem[{{MIPS Instrument And MIPS Instrument Support Teams}(2011)}]{mips_handbook}
{MIPS Instrument And MIPS Instrument Support Teams}. 2011, 100, \dodoi{10.26131/IRSA488}

\bibitem[{{Mizuno} {et~al.}(1990){Mizuno}, {Fukui}, {Iwata}, {Nozawa}, \& {Takano}}]{mizuno1990}
{Mizuno}, A., {Fukui}, Y., {Iwata}, T., {Nozawa}, S., \& {Takano}, T. 1990, \apj, 356, 184, \dodoi{10.1086/168829}

\bibitem[{{Moe} \& {Di Stefano}(2017)}]{moe2017}
{Moe}, M., \& {Di Stefano}, R. 2017, \apjs, 230, 15, \dodoi{10.3847/1538-4365/aa6fb6}

\bibitem[{{Moe} \& {Kratter}(2018)}]{moe2018}
{Moe}, M., \& {Kratter}, K.~M. 2018, \apj, 854, 44, \dodoi{10.3847/1538-4357/aaa6d2}

\bibitem[{{Novak}(2011)}]{Novak2011}
{Novak}, G. 2011, in Astronomical Society of the Pacific Conference Series, Vol. 449, Astronomical Polarimetry 2008: Science from Small to Large Telescopes, ed. P.~{Bastien}, N.~{Manset}, D.~P. {Clemens}, \& N.~{St-Louis}, 50

\bibitem[{{Novak} {et~al.}(1997){Novak}, {Dotson}, {Dowell}, {Goldsmith}, {Hildebrand}, {Platt}, \& {Schleuning}}]{novak1997}
{Novak}, G., {Dotson}, J.~L., {Dowell}, C.~D., {et~al.} 1997, \apj, 487, 320, \dodoi{10.1086/304576}

\bibitem[{{Padoan} \& {Nordlund}(2002)}]{padoan2002}
{Padoan}, P., \& {Nordlund}, {\r{A}}. 2002, \apj, 576, 870, \dodoi{10.1086/341790}

\bibitem[{{Pattle} {et~al.}(2023){Pattle}, {Fissel}, {Tahani}, {Liu}, \& {Ntormousi}}]{pattle2023}
{Pattle}, K., {Fissel}, L., {Tahani}, M., {Liu}, T., \& {Ntormousi}, E. 2023, in Astronomical Society of the Pacific Conference Series, Vol. 534, Protostars and Planets VII, ed. S.~{Inutsuka}, Y.~{Aikawa}, T.~{Muto}, K.~{Tomida}, \& M.~{Tamura}, 193, \dodoi{10.48550/arXiv.2203.11179}

\bibitem[{{Pattle} {et~al.}(2015){Pattle}, {Ward-Thompson}, {Kirk}, {White}, {Drabek-Maunder}, {Buckle}, {Beaulieu}, {Berry}, {Broekhoven-Fiene}, {Currie}, {Fich}, {Hatchell}, {Kirk}, {Jenness}, {Johnstone}, {Mottram}, {Nutter}, {Pineda}, {Quinn}, {Salji}, {Tisi}, {Walker-Smith}, {di Francesco}, {Hogerheijde}, {Andr{\'e}}, {Bastien}, {Bresnahan}, {Butner}, {Chen}, {Chrysostomou}, {Coude}, {Davis}, {Duarte-Cabral}, {Fiege}, {Friberg}, {Friesen}, {Fuller}, {Graves}, {Greaves}, {Gregson}, {Griffin}, {Holland}, {Joncas}, {Knee}, {K{\"o}nyves}, {Mairs}, {Marsh}, {Matthews}, {Moriarty-Schieven}, {Rawlings}, {Richer}, {Robertson}, {Rosolowsky}, {Rumble}, {Sadavoy}, {Spinoglio}, {Thomas}, {Tothill}, {Viti}, {Wouterloot}, {Yates}, \& {Zhu}}]{Pattle2015}
{Pattle}, K., {Ward-Thompson}, D., {Kirk}, J.~M., {et~al.} 2015, \mnras, 450, 1094, \dodoi{10.1093/mnras/stv376}

\bibitem[{{Pattle} {et~al.}(2017){Pattle}, {Ward-Thompson}, {Berry}, {Hatchell}, {Chen}, {Pon}, {Koch}, {Kwon}, {Kim}, {Bastien}, {Cho}, {Coud{\'e}}, {Di Francesco}, {Fuller}, {Furuya}, {Graves}, {Johnstone}, {Kirk}, {Kwon}, {Lee}, {Matthews}, {Mottram}, {Parsons}, {Sadavoy}, {Shinnaga}, {Soam}, {Hasegawa}, {Lai}, {Qiu}, \& {Friberg}}]{Pattle2017}
{Pattle}, K., {Ward-Thompson}, D., {Berry}, D., {et~al.} 2017, \apj, 846, 122, \dodoi{10.3847/1538-4357/aa80e5}

\bibitem[{{Pattle} {et~al.}(2021){Pattle}, {Lai}, {Di Francesco}, {Sadavoy}, {Ward-Thompson}, {Johnstone}, {Hoang}, {Arzoumanian}, {Bastien}, {Bourke}, {Coud{\'e}}, {Doi}, {Eswaraiah}, {Fanciullo}, {Furuya}, {Hwang}, {Hull}, {Kang}, {Kim}, {Kirchschlager}, {Kwon}, {Kwon}, {Lee}, {Liu}, {Redman}, {Soam}, {Tahani}, {Tamura}, \& {Tang}}]{pattle2021}
{Pattle}, K., {Lai}, S.-P., {Di Francesco}, J., {et~al.} 2021, \apj, 907, 88, \dodoi{10.3847/1538-4357/abcc6c}

\bibitem[{{Pattle} {et~al.}(2022){Pattle}, {Lai}, {Sadavoy}, {Coud{\'e}}, {Wolf}, {Furuya}, {Kwon}, {Lee}, \& {Zielinski}}]{Pattle2022}
{Pattle}, K., {Lai}, S.-P., {Sadavoy}, S., {et~al.} 2022, \mnras, 515, 1026, \dodoi{10.1093/mnras/stac1356}

\bibitem[{{Pineda} {et~al.}(2012){Pineda}, {Maury}, {Fuller}, {Testi}, {Garc{\'\i}a-Appadoo}, {Peck}, {Villard}, {Corder}, {van Kempen}, {Turner}, {Tachihara}, \& {Dent}}]{pineda2012}
{Pineda}, J.~E., {Maury}, A.~J., {Fuller}, G.~A., {et~al.} 2012, \aap, 544, L7, \dodoi{10.1051/0004-6361/201219589}

\bibitem[{{Planck Collaboration} {et~al.}(2016){Planck Collaboration}, {Ade}, {Aghanim}, {Alves}, {Arnaud}, {Arzoumanian}, {Ashdown}, {Aumont}, {Baccigalupi}, {Banday}, {Barreiro}, {Bartolo}, {Battaner}, {Benabed}, {Beno{\^\i}t}, {Benoit-L{\'e}vy}, {Bernard}, {Bersanelli}, {Bielewicz}, {Bock}, {Bonavera}, {Bond}, {Borrill}, {Bouchet}, {Boulanger}, {Bracco}, {Burigana}, {Calabrese}, {Cardoso}, {Catalano}, {Chiang}, {Christensen}, {Colombo}, {Combet}, {Couchot}, {Crill}, {Curto}, {Cuttaia}, {Danese}, {Davies}, {Davis}, {de Bernardis}, {de Rosa}, {de Zotti}, {Delabrouille}, {Dickinson}, {Diego}, {Dole}, {Donzelli}, {Dor{\'e}}, {Douspis}, {Ducout}, {Dupac}, {Efstathiou}, {Elsner}, {En{\ss}lin}, {Eriksen}, {Falceta-Gon{\c{c}}alves}, {Falgarone}, {Ferri{\`e}re}, {Finelli}, {Forni}, {Frailis}, {Fraisse}, {Franceschi}, {Frejsel}, {Galeotta}, {Galli}, {Ganga}, {Ghosh}, {Giard}, {Gjerl{\o}w}, {Gonz{\'a}lez-Nuevo}, {G{\'o}rski}, {Gregorio}, {Gruppuso}, {Gudmundsson}, {Guillet}, {Harrison}, {Helou}, {Hennebelle},
  {Henrot-Versill{\'e}}, {Hern{\'a}ndez-Monteagudo}, {Herranz}, {Hildebrandt}, {Hivon}, {Holmes}, {Hornstrup}, {Huffenberger}, {Hurier}, {Jaffe}, {Jaffe}, {Jones}, {Juvela}, {Keih{\"a}nen}, {Keskitalo}, {Kisner}, {Knoche}, {Kunz}, {Kurki-Suonio}, {Lagache}, {Lamarre}, {Lasenby}, {Lattanzi}, {Lawrence}, {Leonardi}, {Levrier}, {Liguori}, {Lilje}, {Linden-V{\o}rnle}, {L{\'o}pez-Caniego}, {Lubin}, {Mac{\'\i}as-P{\'e}rez}, {Maino}, {Mandolesi}, {Mangilli}, {Maris}, {Martin}, {Mart{\'\i}nez-Gonz{\'a}lez}, {Masi}, {Matarrese}, {Melchiorri}, {Mendes}, {Mennella}, {Migliaccio}, {Miville-Desch{\^e}nes}, {Moneti}, {Montier}, {Morgante}, {Mortlock}, {Munshi}, {Murphy}, {Naselsky}, {Nati}, {Netterfield}, {Noviello}, {Novikov}, {Novikov}, {Oppermann}, {Oxborrow}, {Pagano}, {Pajot}, {Paladini}, {Paoletti}, {Pasian}, {Perotto}, {Pettorino}, {Piacentini}, {Piat}, {Pierpaoli}, {Pietrobon}, {Plaszczynski}, {Pointecouteau}, {Polenta}, {Ponthieu}, {Pratt}, {Prunet}, {Puget}, {Rachen}, {Reinecke}, {Remazeilles}, {Renault},
  {Renzi}, {Ristorcelli}, {Rocha}, {Rossetti}, {Roudier}, {Rubi{\~n}o-Mart{\'\i}n}, {Rusholme}, {Sandri}, {Santos}, {Savelainen}, {Savini}, {Scott}, {Soler}, {Stolyarov}, {Sudiwala}, {Sutton}, {Suur-Uski}, {Sygnet}, {Tauber}, {Terenzi}, {Toffolatti}, {Tomasi}, {Tristram}, {Tucci}, {Umana}, {Valenziano}, {Valiviita}, {Van Tent}, {Vielva}, {Villa}, {Wade}, {Wandelt}, {Wehus}, {Ysard}, {Yvon}, \& {Zonca}}]{planckXXXV2016}
{Planck Collaboration}, {Ade}, P.~A.~R., {Aghanim}, N., {et~al.} 2016, \aap, 586, A138, \dodoi{10.1051/0004-6361/201525896}

\bibitem[{{Rao} {et~al.}(2009){Rao}, {Girart}, {Marrone}, {Lai}, \& {Schnee}}]{Rao2009}
{Rao}, R., {Girart}, J.~M., {Marrone}, D.~P., {Lai}, S.-P., \& {Schnee}, S. 2009, \apj, 707, 921, \dodoi{10.1088/0004-637X/707/2/921}

\bibitem[{{Rodr{\'\i}guez} {et~al.}(2005){Rodr{\'\i}guez}, {Loinard}, {D'Alessio}, {Wilner}, \& {Ho}}]{Rodriguez2005}
{Rodr{\'\i}guez}, L.~F., {Loinard}, L., {D'Alessio}, P., {Wilner}, D.~J., \& {Ho}, P. T.~P. 2005, \apjl, 621, L133, \dodoi{10.1086/429223}

\bibitem[{{Sadavoy} {et~al.}(2018){Sadavoy}, {Myers}, {Stephens}, {Tobin}, {Kwon}, {Segura-Cox}, {Henning}, {Commer{\c{c}}on}, \& {Looney}}]{Sadavoy2018}
{Sadavoy}, S.~I., {Myers}, P.~C., {Stephens}, I.~W., {et~al.} 2018, \apj, 869, 115, \dodoi{10.3847/1538-4357/aaef81}

\bibitem[{{Sadavoy} {et~al.}(2019){Sadavoy}, {Stephens}, {Myers}, {Looney}, {Tobin}, {Kwon}, {Commer{\c{c}}on}, {Segura-Cox}, {Henning}, \& {Hennebelle}}]{sadavoy2019}
{Sadavoy}, S.~I., {Stephens}, I.~W., {Myers}, P.~C., {et~al.} 2019, \apjs, 245, 2, \dodoi{10.3847/1538-4365/ab4257}

\bibitem[{S{\'{a}}nchez-Portal {et~al.}(2014)S{\'{a}}nchez-Portal, , Marston, Altieri, Aussel, Feuchtgruber, Klaas, Linz, Lutz, Mer{\'{\i}}n, Müller, Nielbock, Oort, Pilbratt, Schmidt, Stephenson, \& Tuttlebee}]{sanchez_portal2014}
S{\'{a}}nchez-Portal, M., , Marston, A., {et~al.} 2014, Experimental Astronomy, 37, 453, \dodoi{10.1007/s10686-014-9396-z}

\bibitem[{{Scarrott} \& {Warren-Smith}(1989)}]{scarrott1989}
{Scarrott}, S.~M., \& {Warren-Smith}, R.~F. 1989, \mnras, 237, 995, \dodoi{10.1093/mnras/237.4.995}

\bibitem[{{Sch{\"o}ier} {et~al.}(2002){Sch{\"o}ier}, {J{\o}rgensen}, {van Dishoeck}, \& {Blake}}]{schoier2002}
{Sch{\"o}ier}, F.~L., {J{\o}rgensen}, J.~K., {van Dishoeck}, E.~F., \& {Blake}, G.~A. 2002, \aap, 390, 1001, \dodoi{10.1051/0004-6361:20020756}

\bibitem[{{Stark} {et~al.}(2004){Stark}, {Sandell}, {Beck}, {Hogerheijde}, {van Dishoeck}, {van der Wal}, {van der Tak}, {Sch{\"a}fer}, {Melnick}, {Ashby}, \& {de Lange}}]{Stark2004}
{Stark}, R., {Sandell}, G., {Beck}, S.~C., {et~al.} 2004, \apj, 608, 341, \dodoi{10.1086/392492}

\bibitem[{{Stephens} {et~al.}(2022){Stephens}, {Myers}, {Zucker}, {Jackson}, {Andersson}, {Smith}, {Soam}, {Battersby}, {Sanhueza}, {Hogge}, {Smith}, {Novak}, {Sadavoy}, {Pillai}, {Li}, {Looney}, {Sugitani}, {Coud{\'e}}, {Guzm{\'a}n}, {Goodman}, {Kusune}, {Santos}, {Zuckerman}, \& {Encalada}}]{Stephens2022}
{Stephens}, I.~W., {Myers}, P.~C., {Zucker}, C., {et~al.} 2022, \apjl, 926, L6, \dodoi{10.3847/2041-8213/ac4d8f}

\bibitem[{{Tobin} {et~al.}(2016){Tobin}, {Looney}, {Li}, {Chandler}, {Dunham}, {Segura-Cox}, {Sadavoy}, {Melis}, {Harris}, {Kratter}, \& {Perez}}]{tobin2016}
{Tobin}, J.~J., {Looney}, L.~W., {Li}, Z.-Y., {et~al.} 2016, \apj, 818, 73, \dodoi{10.3847/0004-637X/818/1/73}

\bibitem[{{Tobin} {et~al.}(2022){Tobin}, {Offner}, {Kratter}, {Megeath}, {Sheehan}, {Looney}, {Diaz-Rodriguez}, {Osorio}, {Anglada}, {Sadavoy}, {Furlan}, {Segura-Cox}, {Karnath}, {van't Hoff}, {van Dishoeck}, {Li}, {Sharma}, {Stutz}, \& {Tychoniec}}]{tobin2022}
{Tobin}, J.~J., {Offner}, S. S.~R., {Kratter}, K.~M., {et~al.} 2022, \apj, 925, 39, \dodoi{10.3847/1538-4357/ac36d2}

\bibitem[{{Tsukamoto} {et~al.}(2023){Tsukamoto}, {Maury}, {Commercon}, {Alves}, {Cox}, {Sakai}, {Ray}, {Zhao}, \& {Machida}}]{tsukamoto2023}
{Tsukamoto}, Y., {Maury}, A., {Commercon}, B., {et~al.} 2023, in Astronomical Society of the Pacific Conference Series, Vol. 534, Astronomical Society of the Pacific Conference Series, ed. S.~{Inutsuka}, Y.~{Aikawa}, T.~{Muto}, K.~{Tomida}, \& M.~{Tamura}, 317

\bibitem[{Vaillancourt {et~al.}(2007)Vaillancourt, Chuss, Crutcher, Dotson, Dowell, Harper, Hildebrand, Jones, Lazarian, Novak, \& Werner}]{vaillancourt2007}
Vaillancourt, J.~E., Chuss, D.~T., Crutcher, R.~M., {et~al.} 2007, in Infrared Spaceborne Remote Sensing and Instrumentation {XV}, ed. M.~Strojnik-Scholl ({SPIE}), \dodoi{10.1117/12.730922}

\bibitem[{{van der Wiel} {et~al.}(2019){van der Wiel}, {Jacobsen}, {J{\o}rgensen}, {Bourke}, {Kristensen}, {Bjerkeli}, {Murillo}, {Calcutt}, {M{\"u}ller}, {Coutens}, {Drozdovskaya}, {Favre}, \& {Wampfler}}]{vanderwiel2019}
{van der Wiel}, M.~H.~D., {Jacobsen}, S.~K., {J{\o}rgensen}, J.~K., {et~al.} 2019, \aap, 626, A93, \dodoi{10.1051/0004-6361/201833695}

\bibitem[{{Vrba} {et~al.}(1976){Vrba}, {Strom}, \& {Strom}}]{Vrba1976}
{Vrba}, F.~J., {Strom}, S.~E., \& {Strom}, K.~M. 1976, \aj, 81, 958, \dodoi{10.1086/111976}

\bibitem[{{Walker} {et~al.}(1988){Walker}, {Lada}, {Young}, \& {Margulis}}]{walker1988}
{Walker}, C.~K., {Lada}, C.~J., {Young}, E.~T., \& {Margulis}, M. 1988, \apj, 332, 335, \dodoi{10.1086/166659}

\bibitem[{{Wootten}(1989)}]{al1989}
{Wootten}, A. 1989, \apj, 337, 858, \dodoi{10.1086/167156}

\bibitem[{{Wootten} \& {Loren}(1987)}]{al1987}
{Wootten}, A., \& {Loren}, R.~B. 1987, \apj, 317, 220, \dodoi{10.1086/165270}

\bibitem[{Wright {et~al.}(2010)Wright, Eisenhardt, Mainzer, Ressler, Cutri, Jarrett, Kirkpatrick, Padgett, McMillan, Skrutskie, Stanford, Cohen, Walker, Mather, Leisawitz, Gautier, McLean, Benford, Lonsdale, Blain, Mendez, Irace, Duval, Liu, Royer, Heinrichsen, Howard, Shannon, Kendall, Walsh, Larsen, Cardon, Schick, Schwalm, Abid, Fabinsky, Naes, \& Tsai}]{wright2010}
Wright, E.~L., Eisenhardt, P. R.~M., Mainzer, A.~K., {et~al.} 2010, The Astronomical Journal, 140, 1868, \dodoi{10.1088/0004-6256/140/6/1868}

\bibitem[{{Yang} {et~al.}(2016){Yang}, {Li}, {Looney}, \& {Stephens}}]{yang2016}
{Yang}, H., {Li}, Z.-Y., {Looney}, L., \& {Stephens}, I. 2016, \mnras, 456, 2794, \dodoi{10.1093/mnras/stv2633}

\bibitem[{{Yeh} {et~al.}(2008){Yeh}, {Hirano}, {Bourke}, {Ho}, {Lee}, {Ohashi}, \& {Takakuwa}}]{Yeh2008}
{Yeh}, S. C.~C., {Hirano}, N., {Bourke}, T.~L., {et~al.} 2008, \apj, 675, 454, \dodoi{10.1086/524648}

\bibitem[{{Zamponi} {et~al.}(2021){Zamponi}, {Maureira}, {Zhao}, {Liu}, {Ilee}, {Forgan}, \& {Caselli}}]{zamponi2021}
{Zamponi}, J., {Maureira}, M.~J., {Zhao}, B., {et~al.} 2021, \mnras, 508, 2583, \dodoi{10.1093/mnras/stab2657}

\bibitem[{{Zhang} {et~al.}(2014){Zhang}, {Qiu}, {Girart}, {Liu}, {Tang}, {Koch}, {Li}, {Keto}, {Ho}, {Rao}, {Lai}, {Ching}, {Frau}, {Chen}, {Li}, {Padovani}, {Bontemps}, {Csengeri}, \& {Ju{\'a}rez}}]{Zhang2014}
{Zhang}, Q., {Qiu}, K., {Girart}, J.~M., {et~al.} 2014, \apj, 792, 116, \dodoi{10.1088/0004-637X/792/2/116}

\bibitem[{{Zielinski} {et~al.}(2021){Zielinski}, {Wolf}, \& {Brunngr{\"a}ber}}]{Zielinski2021}
{Zielinski}, N., {Wolf}, S., \& {Brunngr{\"a}ber}, R. 2021, \aap, 645, A125, \dodoi{10.1051/0004-6361/202039126}

\end{thebibliography}
